\documentclass[10pt,twocolumn]{article}

\usepackage{caption}
\usepackage{latex8}                              
\usepackage{times}                               

\usepackage[sf]{subfigure}
\usepackage{fullpage}
\usepackage[dvips]{graphicx}
\usepackage[]{rotating}
\usepackage{textcomp}
\usepackage{ifthen}
\usepackage{setspace}

\let \Mysubcapsize \subcapsize
\renewcommand{\subcapsize}{\Mysubcapsize\ninehv}

\let \Mysection \section
\newcommand{\Mysectionstar}[1]{\Mysection*{#1}
\vspace{-10pt}}

\renewcommand{\section}[1]{\vspace{-7pt}
\ifthenelse{\equal{#1}{*}}
{\Mysectionstar}
{\Mysection{#1}
\vspace{-10pt}}}

\let \MySubSection \SubSection
\renewcommand{\SubSection}[1]{\vspace{-7pt}
\MySubSection{#1}
\vspace{-10pt}}

\let \Mysubsubsection \subsubsection
\renewcommand{\subsubsection}[1]{\vspace{-8pt}
\Mysubsubsection{#1}
\vspace{-6pt}}

\graphicspath{{graphs/}}

\begin{document}


\title{
  Efficient and Scalable Barrier over Quadrics and Myrinet with \\
  a New NIC-Based Collective Message Passing Protocol
  \thanks{This research is 
  supported in part by a DOE grant \#DE-FC02-01ER25506,
  NSF Grants \#EIA-9986052 and \#CCR-0204429, a grant from 
  Los Alamos National Laboratory, and also in part by 
  U.S. Department of Energy, under Contract W-31-109-Eng-38.}
}

\author{
  \begin{tabular}{cccc} 
  Weikuan Yu$^\dag$ & Darius Buntinas$^\ddagger$ 
  & Rich L. Graham$^{\S}$ & Dhabaleswar K. Panda$^\dag$ \\ 
  \end{tabular}
  \\ \\
  \begin{tabular}{ccc} 
  {\normalsize Network-Based Computing Lab}$^\dag$ &
  {\normalsize Argonne National Laboratory}$^\ddagger$ &
  {\normalsize Los Alamos National Laboratory}$^{\S}$  \\
  {\normalsize Dept.~of Computer and Info.~Science} &
  {\normalsize Mathematics and Computer Science}  &
  {\normalsize Advanced Computing Laboratory}  \\
  {\normalsize The Ohio State University} &
  {\normalsize Argonne, IL 60439}  &
  {\normalsize Los Alamos, NM 87545} \\
  {\normalsize \{yuw,panda\}@cis.ohio-state.edu} &
  {\normalsize buntinas@mcs.anl.gov} &
  {\normalsize rlgraham@lanl.gov} \\
  \end{tabular}
}

\date{}
\maketitle
\thispagestyle{empty}

\begin{abstract}
Modern interconnects often have 
programmable processors in the network interface that 
can be utilized to offload communication processing from host CPU.
In this paper, we explore different schemes to support
collective operations at the network interface and 
propose a new collective protocol. 
With barrier as an initial case study, we have demontrated that 
much of the communication processing can be greatly simplified with 
this collective protocol.
Accordingly, 
we have designed and implemented 
efficient and scalable NIC-based barrier operations
over two high performance interconnects, Quadrics and Myrinet.

Our evaluation shows that, over a Quadrics cluster of 8 nodes with 
ELan3 Network, the NIC-based barrier operation 
achieves a barrier latency of only 5.60$\mu$s. This result is a 2.48 
factor of improvement over the Elanlib tree-based barrier operation.
Over a Myrinet cluster of 8 nodes with LANai-XP NIC cards, 
a barrier latency of 14.20$\mu$s over 8 nodes is achieved.
This is a 2.64 factor of improvement over the host-based barrier algorithm.
Furthermore, 
an analytical model developed for the proposed scheme indicates
that a NIC-based barrier operation on a 1024-node cluster can be
performed with only 22.13$\mu$s latency over Quadrics and
with 38.94$\mu$s latency over Myrinet.
These results indicate the potential for developing high performance
communication subsystems for next generation clusters.

\end{abstract}


\Section{Introduction}
\label{Sec:intro}
Barrier is a commonly used collective operation 
in parallel and distributed programs. 
Message passing standards, such as MPI~\cite{MPI:mpi_2_96},
often have the barrier operation included as 
a part of their specifications.  
In the function MPI\_Barrier(), 
while processes are performing the barrier communication and 
waiting for its completion, no other computation can be performed.
So it is important to minimize the amount of time spent 
on waiting for the barrier.  The efficiency of barrier
also affects the granularity of a parallel application.  
To support fine-grained parallel applications,
an efficient barrier primitive must be provided.
Some modern interconnects, such as
QsNet~\cite{Petrini:Quadrics_micro_02_journal} and
InfiniBand~\cite{Eddington:infinibridge_02}, 
provide hardware broadcast primitives that can be utilized to 
support an efficient barrier operation. 
However, hardware broadcast primitives often have their limitations. 
For example, Quadrics hardware broadcast
requires that all the processes are located on 
a contiguous set of nodes and also well synchronized during its
computation to achieve high performance barrier operations;
InfiniBand hardware broadcast is not reliable.
Other interconnects, such as Myrinet, 
do not have hardware broadcast and provide unicast communication 
along point-to-point links. 
Thus, a general barrier operation is often implemented 
on top of point-to-point communication.

Earlier research has been done to use programmable processors 
to support efficient collective 
operations~\cite{Yuw:High_icpp_03_bcast,Moody:Scalable_sc_03_reduce, 
Buntinas:barrier_01}.
Among them, Buntinas et.~al.~\cite{Buntinas:barrier_01} has explored 
NIC-based barrier over Myrinet/GM. 
In that study, the NIC takes an active role in detecting 
arrived barrier messages and triggering the next barrier messages.
This greatly reduces round-trip PCI bus traffic and host CPU involvement 
in a barrier operation, thereby improving the barrier latency.
However, much of the communication processing for barrier messages
is still implemented on top of the NIC's point-to-point 
communication processing.  
The benefits of NIC-based barrier have been exposed, 
but only to a certain extent.
And the scheme has not been generalized to expose the benefits of NIC
programmability over other networks, for example, Quadrics.
So it remains an open challenge to gain more insights into 
the related communication processing and propose an efficient, 
and generally applicable scheme in order to provide maximum benefits 
to NIC-based barrier operations.

In this paper, we take on this challenge.
We start with discussing the characteristics of NIC-based barrier operations.
We then examine the communication processing tasks
for point-to-point operations, 
including queuing, bookkeeping, packetizing and assembly, 
flow control and error control, etc.
Many of these tasks are redundant for collective operations.
We then propose a novel NIC-based collective protocol
which performs queuing, bookkeeping, packetizing and error control tasks
in a collective manner and eliminates the redundancy wherever possible.
With barrier as an initial case study, we have demonstrated that 
much of the communication processing can be greatly simplified.
Accordingly, the proposed scheme is implemented over Myrinet. 
Furthermore, a similar NIC-based barrier is implemented over Quadrics.

Our evaluation has shown that, over a Quadrics cluster of 8 nodes with 
ELan3 Network, the NIC-based barrier operation 
achieves a barrier latency of 5.60$\mu$s. This result is a 2.48 
factor of improvement over the Elanlib tree-based barrier operation.
Over a Myrinet cluster of 8 nodes with LANai-XP NIC cards, 
a barrier latency of 14.20$\mu$s over 8 nodes is achieved.
This is a 2.64 factor of improvement over the host-based barrier algorithm.
Our evaluation has also shown that, over a 16-node Myrinet cluster 
with LANai 9.1 cards, the NIC-based barrier operation achieves
a barrier latency of 25.72us, 
a 3.38 factor of improvement compared to the host-based algorithm.
Furthermore, our analytical model suggests that 
NIC-based barrier operations 
could achieve a latency of only 22.13$\mu$s and 38.94$\mu$s,
respectively over a 1024-node Quadrics and Myrinet cluster.

The rest of the paper is structured as follows. 
In the next section, we explore different NIC-based barrier algorithms 
and describe the motivation for the NIC-based barrier with 
a separate collective protocol.
Following that, we describe in detail the design issues of 
the barrier operations in the proposed NIC-based collective protocol.
Then in Section~\ref{Sec:myrinet-qsnet}
we give an overview of Quadrics and Myrinet.
An overview of the barrier algorithms considered for our implementation
is provided in Section~\ref{Sec:algorithms}.
In Sections~\ref{Sec:barrier-myrinet} and~\ref{Sec:barrier-quadrics},
we describe our NIC-based barrier operations over Myrinet and Quadrics.
The performance results are provided
in Section~\ref{Sec:results}. 
Finally, we conclude the paper in Section~\ref{Sec:discussion}.

\Section{Motivation}
\label{Sec:motivation}

\begin{figure*}[htb] 
  \centering
  \includegraphics[width=0.80\textwidth]{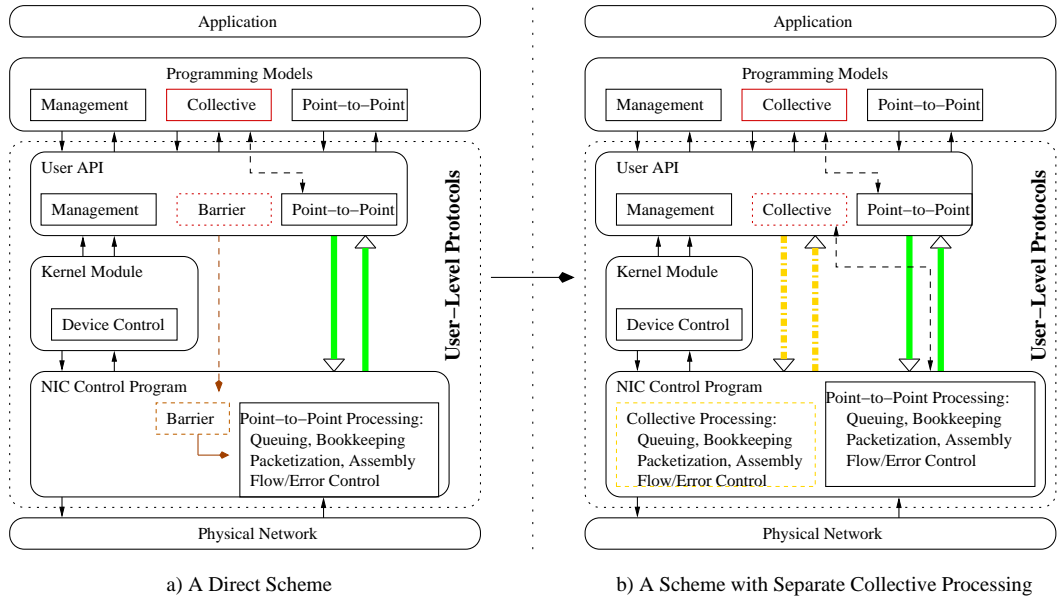}
  \vspace{-12pt}
  \caption{Different Schemes to Support NIC-based 
           Collective(Barrier) Operations}
  \label{fig:prot}
  \vspace{1pc}
\end{figure*}

In this section, we describe general ideas 
of previous research~\cite{Buntinas:barrier_01, 
Buntinas:Performance_cac_01_barrier}
on NIC-based barrier operations over point-to-point communication. 
In addition, we explore different ideas to support collective
communication and the motivation for NIC-based barrier operations
with a separate collective protocol.

\SubSection{Previous Research on NIC-Based Barrier}
\label{SubSec:previous}

Buntinas et.~al.~\cite{Buntinas:barrier_01, 
Buntinas:Performance_cac_01_barrier}
have studied the benefits of off-loading barrier operation to 
the Myrinet Control Program (MCP). 
With the previous NIC-based barrier scheme,
the NIC takes an active role in performing the barrier operation.
Host CPU is not involved in the intermediate steps of a barrier operation.
The number of round-trip messages across the PCI bus is reduced. 
However, further investigation into this implementation reveals that 
it builds the NIC-based barrier operation simply on top of 
the point-to-point communication protocol running on the NIC. 
The left diagram in Fig.~\ref{fig:prot} shows how the barrier
implementation fits into a user-level protocol (in this case, MCP).
With this approach, much of the communication processing is redundant 
for the nature of barrier operations.
It still remains to be examined how much redundant processing is done.
Likewise, it is not analyzed how much benefits there are
if one can eliminate the redundancy with a separate collective protocol.

\SubSection{The Point-to-Point Communication Protocol at the NIC}
\label{SubSec:pt2pt}
In a NIC control program for a general user-level protocol, 
this processing can be classified into the following categories of
tasks: request queuing, request bookkeeping, data packetization,
data assembly, flow control and error control.
An overview to the communication processing performed by
the Myrinet Control Program is provided in Section~\ref{SubSec:myrinet}.
The communication tasks are usually well-tuned for point-to-point 
communication. But to achieve high performance NIC-based 
collective operations, much of these tasks can be done in a collective 
manner.  This can lead to simplified and reduced processing.
Thus a separate communication protocol for the NIC-based collective 
operations is needed to maximize the benefits.

\SubSection{Where to Provide Support for Collective Communication?}
\label{SubSec:where}
The performance of the resulting
collective operations are often limited by the underlying user-level
protocols.  If the user-level protocols only provide 
point-to-point communication semantics,
the programming models have to lay their collective support
on top of that.  The resulting performance may not be ideal.
The NIC-based collective operations can help expose 
the best performance from the underlying network to these developers.
However, as shown in the left diagram of Fig.~\ref{fig:prot},
the earlier NIC-based barrier implementation intercepts the requests 
for the barrier operations and  directly delivers the barrier messages.
No efforts have been put to examine how the communication processing tasks
are undertaken by the NICs for these regular messages
and how to reduce them for barrier operations.
Thus this direct scheme of offloading the barrier
operation does not achieve maximum benefits. 

\Section{A Proposed Scheme to Support NIC-based Barrier Operations}
\label{Sec:proposed}

In this section, 
we propose a novel scheme with a NIC-based collective protocol 
to eliminate the redundancy described in the last section.
Then with barrier operations as the focus in this paper, 
we describe how the benefits of NIC-based barrier operations 
can be maximized with this scheme.
The associated design issues are also discussed.

As shown in the right diagram of Fig.~\ref{fig:prot}, 
a separate protocol at the NIC is proposed to perform
the communication processing tasks related to collective operations.
To provide an interface to the procotol, a set of
API's for collective operations can be provided at the user-level.
Then the support for these collective operations can be implemented
at the NIC. If there is any collective operation that cannot
be supported efficiently by the NIC, 
its implementation can still be laid over point-to-point protocols.
Basically, our scheme aims to provide a protocol 
that collectively performs the message passing tasks necessary 
for collective operations. 
For each collective operation,
the critical step is to identify the tasks
that can be more efficiently put into the collective protocol. 
In the case of a NIC-based barrier operation, 
we have identified the following tasks 
that need to be included in the NIC-based collective protocol.

\begin{description}
\item[Queuing] \ \\
In a parallel system, a NIC must handle multiple communication requests 
to a peer NIC and also requests to multiple different peer NICs.
Each request must go through multiple queues and 
be scheduled before the message can be transmitted. 
Thus for a barrier, the arrived message may not immediately 
lead to the transmission of the next message until the corresponding request 
gets its turn in the relevant queues.
This imposes unnecessary delays into the barrier operations.
If we can provide a separate queue for a particular process group,
its barrier messages can skip other queues and get transmitted in
a much faster manner. 

\item[Packetization and Assembly] \ \\
The sender NIC must packetize the large messages and
allocate a send buffer for each of the packet.
For that the NIC has to wait for a send buffer to become available
and fill up the packet with data before the messaging takes
place.  Since all the information a barrier message needs to 
carry along is an integer, if one can utilize a dedicated send buffer 
for the barrier messages, all these unnecessary waiting 
for a send buffer can also be eliminated.
At the receiver side, the received barrier message 
also does not need to go through the queues for data assembly, etc.

\item[Bookkeeping] \ \\
For each outstanding messaging request,
the NIC must perform bookkeeping functions to keep track of 
its status of every packet transmitted on its behalf.
This is rather inefficient for a barrier operation,
since there is no data transmission involved.
One can just provide a bit vector to record whether 
all the messages for a barrier operation are completed or not.

\item[Flow/Error Control] \ \\
Depending on the reliability feature of the underlying network,
the NIC control program may also need to provide flow control 
and/or error control functions to ensure reliability.
The error control for point-to-point messages is usually 
implemented with a form of timeout/retransmission. 
Acknowledgments are returned from the receivers to the senders.
The NIC-based barrier also provides opportunities
to have an efficient and simplified error control.
For example, we can eliminate all the acknowledgments and 
provide reliability with a receiver-driven retransmission approach. 
When a barrier operation fails to complete due to 
the missing of some barrier messages, 
NACKs can be sent to the corresponding senders.
Thus this reduces the number of actual barrier messages 
by half and can speed up the barrier operation.
\end{description} 

\Section{Overview of Quadrics and Myrinet}
\label{Sec:myrinet-qsnet}
In this section, we describe some background information on two
interconnects that provide programmable NIC processors,
Quadrics and Myrinet. Quadrics provides hardware-level
reliable message passing, while Myrinet does not. 
The message passing reliability is left to the communication protocol.
Designing an efficient reliability scheme is then critical to 
the performance of the communication protocol over such a network.

\SubSection{Quadrics and Elanlib}
\label{SubSec:quadrics}
Quadrics network (QsNet)~\cite{Petrini:Quadrics_micro_02_journal} 
provides low-latency, high-bandwidth
communication with its two building blocks: a programmable Elan
network interface card and the Elite switch, which are
interconnected in a fat-tree topology. 
QsNet-II~\cite{Beecroft:qsnet_02}
has been released recently, but in the scope of this paper, 
a Quadrics interconnect with Elan3 network interface cards.
We are planning to extend similar studies to QsNet-II  
once such an system becomes available to us.

{\bf QsNet Programming Library} -- 
QsNet provides the Elan and Elan3 libraries as the interface for  
its Elan3 network~\cite{Quadrics:Web}. 
At the Elan3 level, a process in a parallel job is allocated a virtual
process id (VPID).  Interprocess communication is supported by 
an efficient model: remote direct memory access (RDMA). 
Elan3lib also provides a very useful {\em chained event}
mechanism, which allows one RDMA descriptor to be triggered upon 
the completion of another RDMA descriptor.
A higher-level programming library, Elanlib, extends Elan3lib with 
point-to-point, tagged message passing primitives (called Tagged Message
Ports or Tports) and support for collective operations.

{\bf Barrier in Elanlib} -- 
Elanlib provides two barrier functions,
elan\_gsync() and elan\_hgsync(). The latter takes advantages of the
hardware broadcast primitive and provides a very efficient and scalable
barrier operation~\cite{Petrini:Hardware_nca_01_coll}.
However, it requires that 
the calling processes are 
well synchronized in their stages of 
computation~\cite{Petrini:Hardware_nca_01_coll}.
Otherwise, it falls back on the elan\_gsync() to complete the barrier
with a tree-based gather-broadcast algorithm.

\SubSection{Myrinet and GM}
\label{SubSec:myrinet}
Myrinet is a high-speed interconnect technology 
using wormhole-routed crossbar switches to connect all the NICs. 
GM is a user-level communication protocol that runs over the 
Myrinet~\cite{Boden:Myrinet_95} and provides a reliable
ordered delivery of packets with low latency and high bandwidth.
The basic send/receive operation works as follows.

{\bf Sending a Message} -- To send a message,
a user application generates a send descriptor, referred
to as a {\em send event} in GM, to the NIC.
The NIC translates the event to 
a {\em send token} (a form of send descriptor that NIC uses), 
and appends it to the send queue for the desired destination. 
With outstanding send tokens to multiple destinations, the NIC
processes the tokens to different destinations in a round-robin manner. 
To send a message for a token, the NIC also has to wait for 
the availability of a send packet, i.e., the send buffer to 
accommodate the data. Then the data is DMAed from the host buffer 
into the {\em send packet} and injected into the network.
The NIC keeps a {\em send record}
of the sequence number and the time for each packet it has sent.
If the acknowledgment is not received within the 
timeout period, the sender will retransmit the packet.  
When all the send records 
are acknowledged, the NIC will pass the send token back to the host. 

{\bf Receiving a Message} -- To receive a message, 
the host provides some registered memory as the receive buffer 
by preposting a receive descriptor. 
A posted receive descriptor is translated
into a {\em receive token} by the NIC. 
When the NIC receives a packet, it checks the sequence number.
An unexpected packet is dropped immediately.
For an expected packet, the NIC locates a receive token,
DMAs the packet data into the host memory, and then acknowledges the sender.
When all the packets for a message have been received,
the NIC will also generate a {\em receive event} to 
the host process for it to detect the arrived message.

\Section{Overview of Barrier Algorithms}
\label{Sec:algorithms}

In this section, we give a brief introduction 
to general barrier algorithms.
Note that we focus on the algorithms for the barrier operation 
on top of point-to-point communication. 
Barrier operations on top of hardware broadcast have been studied 
in~\cite{Petrini:Hardware_nca_01_coll} 
and~\cite{Kinis:Fast_euro_20003_barrier}.

\SubSection{General Algorithms}
\label{SubSec:algorithms}
Without using hardware barrier/broadcast primitives,
a barrier operation typically requires the exchange of
multiple point-to-point messages between processes.
Typically it is implemented by one of the following three algorithms:
gather-broadcast~\cite{McKinley:Survey_94_collective},
pairwise-exchange~\cite{Gropp:High_96_mpich}
and dissemination~\cite{Kinis:Fast_euro_20003_barrier}.

\begin{figure}[htb]
  \centering
  \includegraphics[height=1.2in]{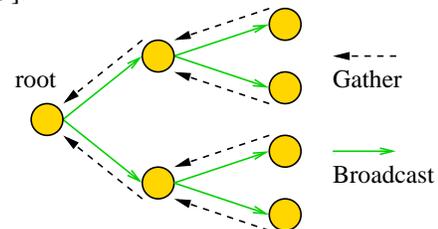}
  \protect\vspace*{-1pc}
  \caption{Gather-Broadcast}
  \label{fig:gather_bcast}
\end{figure}
\begin{figure}[htb]
  \centering
  \protect\vspace*{.5pc}
  \includegraphics[height=1.2in]{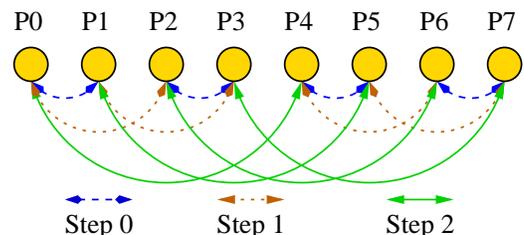}
  \protect\vspace*{-1pc}
  \caption{Pairwise-Exchange}
  \label{fig:pairwise}
\end{figure}

{\bf Gather-Broadcast} -- As shown in Fig.~\ref{fig:gather_bcast}, 
processes involved in a barrier form a tree-based topology. 
All the barrier messages are propagated up the tree and 
combined to the root, which in turn broadcasts a message down 
the tree to have other processes exit
the barrier.  For a group of $N$ participating nodes,
this algorithm takes ($2 \log_d N$) steps, 
where $d$ is the degree of the tree.

\begin{figure}[htb]
  \centering
  \protect\vspace*{.5pc}
  \includegraphics[width=0.95\columnwidth]{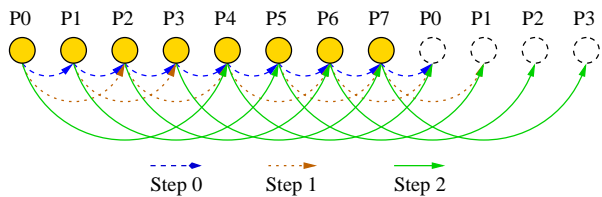}
  \protect\vspace*{-1pc}
  \caption{Dissemination}
  \label{fig:dissemination}
\end{figure}

{\bf Pairwise-Exchange} -- 
This is a recursive doubling algorithm used in the popular 
MPICH~\cite{Gropp:High_96_mpich} distribution. 
As shown in Fig.~\ref{fig:pairwise},
at step m, process i and j, where $j = i \wedge 2^m$,
are paired up and exchange messages. 
For a group of $N$ participating nodes,
this algorithm takes $\log_2 N$ steps, 
when $N$ is a power of two.  If $N$ is not a power of two, 
two additional steps needs to be performed.
Let M be the largest power of 2 and less than N.
At the very beginning, process i sends a message to processes j, 
where $i \ge {2^m}$ and $j = i - {2^m}$.
Then the low ranked M processes perform pairwise exchange for the
barrier. At the very end, process j notifies process i to exit the barrier.
This algorithm takes ($\lfloor\log_2 N\rfloor + 2$) steps
for non-power of two number of nodes.

{\bf Dissemination} -- 
This dissemination algorithm is also
described in~\cite{Crummey:Algorithms_1991_ACM_sync}.
As shown in Fig.~\ref{fig:dissemination},
in step $m$, process $i$
sends a barrier message to process $j$, where $j = (i + {2^m})mod N$.
Essentially, barrier messages are disseminated around processes so that 
each process is able to collect the barrier information
from its left ${2^{m+1}}$ processes by step m.
This algorithm takes $\lceil\log_2 N\rceil$ steps, 
irrespective of whether N is a power of two or not.

\SubSection{Choosing the Right Algorithm}
\label{SubSec:choice}
From the earlier description, it is clear that 
the gather-broadcast algorithm requires more steps 
for a barrier operation. 
Buntinas et.~al.~\cite{Buntinas:barrier_01,
Buntinas:Performance_cac_01_barrier} also have found that the
pairwise-exchange algorithm generally performs better 
than the gather-broadcast algorithm.
Thus for the proposed NIC-based barrier in this paper,
we have chosen to implement and compare the pairwise-exchange 
and dissemination algorithms.

\Section{Implementation of the Proposed NIC-Based Barrier over Myrinet}
\label{Sec:barrier-myrinet}

In this section, we describe the NIC-based barrier 
over Myrinet/GM.
We have explored many of the challenging issues in our earlier work
with GM-1.2.3~\cite{Buntinas:barrier_01,
Buntinas:Performance_cac_01_barrier}.
As having discussed in Section~\ref{Sec:proposed}, 
we choose to create a separated protocol to process the barrier messages.
We believe that reimplementing the previous work over GM-2.0.3 would
lead to the same amount of relative improvement since the NIC-based
barrier is mainly dependent on the number of messages and processing
steps to be performed. 
Solutions from the earlier work for some of the challenges
have been incorporated into this new protocol. 
Other challenging issues related to 
the new barrier protocol are described in this section.

\SubSection{Queuing the Barrier Operations} 
\label{SubSec:queuing}
As described in Section~\ref{SubSec:myrinet}, 
MCP processes the send tokens to different destinations 
in a round robin fashion. Send tokens to the same destination
are processed in a FIFO manner. 
So the send tokens for barrier operations
must go through multiple queues before their messages can be transmitted.
This is enforced to the initial barrier message 
(e.g., in Step 1 of the pairwise-exchange algorithm) 
and also the barrier message that needs to be transmitted immediately
when an earlier barrier message arrives.
It is rather inefficient to have the NIC-based barrier operations
put up with so much waiting.
We created a separate queue for each group of processes,
and enqueued only one send token for every barrier operation.
Then the barrier messages do not have to go through 
the queues for multiple destinations. With this approach,
the send token for the current barrier operation is always 
located at the front of its queue. 
Both the initial barrier message
and the ones that need to be triggered later no longer need to
go through the queues for the corresponding destinations.

\SubSection{Packetizing the Barrier Messages}
\label{SubSec:packeting} 
Within the Myrinet Control Program, to send any message, 
the sender NIC must wait for a send packet to become
available and fill up the packet with data.
So to complete a barrier operation, it is inevitable 
for the sender NIC to go through multiple rounds of 
allocating, filling up and releasing the send packets.
Since all the information a barrier message needs to carry along 
is an integer, it is much more efficient if a static send packet 
can be utilized to transmit this integer and avoid going through 
multiple rounds of claiming/releasing the send packets.

This static send packet can be very small since it only carries an integer. 
One can allocate an additional short send packet 
for each group of processes.
However, there is a static send packet to each peer NIC in MCP,
which is used for fast transmission of ACKs.
We pad this static packet with an extra integer and 
utilize it in our implementation.
With this approach, all the packetizing
(including packets claiming and releasing) needed for transmitting
regular messages is avoided for the barrier messages.

\SubSection{Bookkeeping and Error Control for Barrier Messages}
\label{SubSec:reliability} 

The Myrinet Control Program provides bookkeeping 
and error control for each packet that has been transmitted.
This is to ensure the reliable delivery of packets.
One acknowledgment must be returned by the receiver in order for 
the sender to release the bookkeeping entries, 
i.e., a send record in MCP.
When a sender NIC fails to receive the ACK within a timeout period 
specified in the send record, it retransmits the packet.
Besides creating multiple send records and keeping track of them,
this also generates twice as many packets as the number of barrier messages.
It is desirable to design a better way
to provide the bookkeeping and error control for the barrier operations 
based on its collective nature.

For the bookkeeping purpose, we create only a send record 
for a barrier operation.  Within the send record, 
a bit vector is provided to keep track of the list of barrier messages.
 When the barrier operation starts,
a time-stamp is also created along with the send record.
In addition, an approach called receiver-driven retransmission 
is provided to ensure reliable delivery of barrier messages.
The receiver NICs of the barrier messages no longer need to return
acknowledgments to the sender NICs.
If any of the expected barrier messages is not received within 
the timeout period, a NACK will be generated from the receiver NIC to 
the corresponding sender NIC. The sender NIC will
then retransmit the barrier message. 
Taken together, these enhancements ensure the reliable delivery
with the minimal possible overhead
and also reduce the number of total packets by half
compared to the reliability scheme for the regular messages.
Thus, it promises a more efficient solution for barrier operation.  

\Section{Implementation of the Proposed NIC-Based Barrier over Quadrics}
\label{Sec:barrier-quadrics}

In this section, we describe the NIC-based barrier over Quadrics.
Quadrics provides salient mechanisms to program the NIC to 
support collective operations~\cite{Moody:Scalable_sc_03_reduce}, 
e.g., threads running in the NIC or 
chained RDMA descriptors. Thus it is rather convenient to implement
NIC-based barrier operation over Quadrics.

Since a barrier operation typically involves no data transfer,
all messages communicated between processes just serve as a form of
notification, indicating that the corresponding processes 
have reached the barrier.  Over Quadrics/Elan, RDMA operations
with no data transfer can be utilized to fire a remote event, 
which serves as a kind of notification to the remote process. 
Although Elan threads can be created and executed by the thread processor
to process the events and chain RDMA operations together, 
an extra thread does increase the processing load to the Elan NIC.
With either pairwise-exchange or dissemination algorithm, all that needed
is to chain the multiple RDMA operations together to 
support a NIC-based barrier. 

We have chosen not to set up an additional thread to support NIC-based
barrier, and instead, set up a list of chained RDMA descriptors at the NIC
from user-level.  The RDMA operations are triggered only upon the arrival of 
a remote event except the very first RDMA operation, which the host process
triggers to initiate a barrier operation. The completion of the very last
RDMA operation will trigger a local event to the host process and
signify the completion of the barrier.

\Section{Performance Evaluation}
\label{Sec:results}
In this section, we describe the performance evaluation of 
our implementation.  The experiments were conducted on two clusters.
One is a 16-node cluster of quad-SMP 700 MHz Pentium-III, 
each equipped with 1GB DRAM and 66MHz/64bit PCI bus.
This cluster is connected with 
both a Myrinet 2000 network and a QsNet/Elan3 network (with only 8 nodes).
The Myrinet NICs have 133MHz LANai 9.1 processors and 2MB SRAM.
The QsNet network consists of a dimension two, quaternary fat tree switch, 
Elite-16, and Elan3 QM-400 cards.
The other system is a cluster of 8-node SuperMicro SUPER P4DL6,
each with dual Intel Xeon 2.4GHz processors, 512MB DRAM,
PCI-X 133MHz/64-bit bus. This cluster is only connected with Myrinet
2000 network and NICs with 225MHz LANai-XP processors and 2MB SRAM.
Our NIC-based implementation over Myrinet is based on GM-2.0.3.
The NIC-based implementation over Quadrics is based on 5.2.7 quadrics
release and Elanlib-1.4.3-2.

\newlength{\myfigheight}
\setlength{\myfigheight}{1.4in}

\SubSection{NIC-Based Barrier over Myrinet}
\label{SubSec:myrinet-results}

\begin{figure}[htb]
    \centering
    \includegraphics[height=\myfigheight]{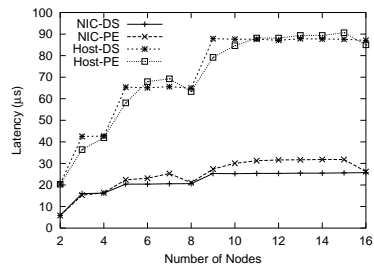}
    \vspace*{-1pc}
    \caption{Performance Evaluation of NIC-based and 
	     Host-Based Barrier Operations with Myrinet LANai-9.1 Cards
             on a 16-node 700MHz cluster}
    \label{fig:barrier_91}
\end{figure}
\begin{figure}[htb]
    \centering
    \includegraphics[height=\myfigheight]{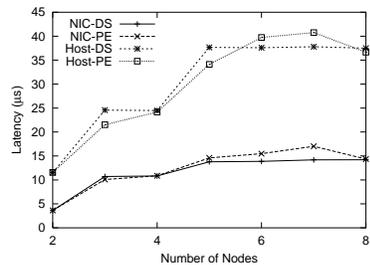}
    \vspace*{-1em}
    \caption{Performance Evaluation of NIC-based and 
	     Host-Based Barrier Operations with Myrinet LANai-XP Cards
             on an 8-node 2.4GHz cluster}
    \label{fig:barrier_xp}
\end{figure}

We tested the latency of our NIC-based barrier operations and
compared it to the host-based barrier operations. 
Our tests were performed by having the processes
execute consecutive barrier operations.
To avoid any possible impact from the network topology and the
allocation of nodes, our tests were performed with 
random permutation of the nodes. We observed only negligible
variations in the performance results.
The first 100 iterations were used to warm up the nodes. 
Then the average for the next 10,000 iterations was taken as the latency.
We compared the performance for both the pairwise-exchange 
and dissemination algorithms. 

Fig.~\ref{fig:barrier_91} 
shows the barrier latencies of NIC-based
and host-based barriers for both algorithms over the 16-node quad-700MHz
cluster with LANai 9.1 cards.
With either pairwise-exchange (PE) or dissemination (DS) algorithm,
the NIC-based barrier operations reduce the barrier latency,
compared to the host-based barrier operations.
The pairwise-exchange algorithm tends to have a larger latency 
over non-power of two number of nodes for the extra step it takes.
Over this 16-node cluster, 
a barrier latency of 25.72$\mu$s is achieved with both algorithms.
This is a 3.38 factor of improvement over host-based barrier
operations.  
Using the direct NIC-based barrier scheme on the same cluster, 
our earlier implementation~\cite{Buntinas:barrier_01,
Buntinas:Performance_cac_01_barrier},
achieved 1.86 factor of improvement using LANai 7.2 cards.
The earlier work was done over GM-1.2.3 and not
maintained as new versions of GM are released. 
We believe that the same amount of relative improvement (1.86) would 
have been achieved if the previous work was 
reimplemented over GM-2.0.3 since the NIC-base
barrier is mainly dependent on the number of messages and processing
steps to be performed.
Although, direct comparisons are not available,
the difference in the improvement factors over the common denominator
(host-based barrier operations) suggests that 
our new scheme provides a large amount of relative benefits.

Fig.~\ref{fig:barrier_xp} 
shows the barrier latencies of NIC-based
and host-based barriers for both algorithms over the eight-node 2.4GHz
Xeon cluster with LANai-XP cards.
Similarly, the NIC-based barrier operation reduces the barrier latency
compared to the host-based barrier operation.
Over this eight node cluster, 
a barrier latency of 14.20$\mu$s is achieved with both algorithms.
This is a 2.64 factor of improvement over the host-based implementation.
The reason that the factor of improvement becomes smaller on this cluster
is because this cluster has a much larger ratio of host CPU speed to
NIC CPU speed and also a faster PCI-X bus. Thus the benefits 
from the reduced host involvement and I/O bus traffic are 
smaller.

\SubSection{NIC-Based Barrier over Quadrics}
\label{SubSec:quadrics-results}

Over an eight-node Quadrics/Elan3 cluster, 
we tested the latency of our NIC-based barrier operations 
and compared them to the elan\_hgsync() function provided in Elanlib,
The performance of elan\_hgsync() is tested 
with hardware broadcast either enabled or disabled.
Our tests were performed by having the processes
perform consecutive barrier operations.
The first 100 iterations were used to warm up the nodes. 
Then the average for the next 10,000 iterations was taken as the latency.
The performance for both the pairwise-exchange 
and dissemination algorithms are compared to the elan\_hgsync()
operation.

\begin{figure}[htb]
  \centering%
    \includegraphics[height=\myfigheight]{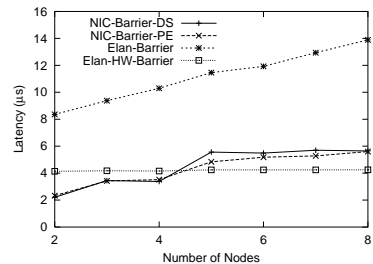}
    \vspace*{-1em}
    \caption{Performance Comparisons of 
             Barrier implementations over Quadrics/Elan3 
             on an 8-node 700MHz cluster}
    \label{fig:barrier_qsnet}
\end{figure}

Fig.~\ref{fig:barrier_qsnet} 
shows the barrier latencies of NIC-based barrier operations
(shown as NIC-Barrier-DS and NIC-Barrier-PE in the figure).
The hardware barrier, elan\_hgsync() achieves a barrier latency of 4.20$\mu$s.
For a small number of nodes, the hardware barrier performs worse than
the NIC-based barrier operation. This is because the hardware barrier is
implemented with an atomic test-and-set operation down the NIC, which
requires a higher number of network transactions.
For a large number of nodes, the hardware barrier performs better
but it requires that the involving processes be well synchronized. 
This is hardly the case for parallel programs over large size clusters. 
Compared to tree-based barrier operation elan\_gsync(),
our NIC-based barrier operation has a much reduced barrier latency.
Note here that the hardware broadcast primitive is disabled for a purely
tree-based barrier with elan\_gsync().
With non-power of two number of nodes,
the pairwise-exchange algorithm performs better than 
the dissemination algorithm over Quadrics.
This is because Quadrics Elan cards is very efficient in 
coping with the hot-spot RDMA operations~\cite{Liuj:Micro_hoti_03}, 
which reduces the effects of the steps for registering and releasing
non-power two processes in a barrier operation.
Over this eight node cluster, 
a barrier latency of 5.60$\mu$s is achieved with both algorithms.
This is a 2.48 factor of improvement over elan\_hgsync() when
the hardware broadcast is not available.

\begin{figure}[htb]
  \subfigure [Over 700MHz Quadrics-Elan3 Cluster] {
    \includegraphics[width=0.47\columnwidth]{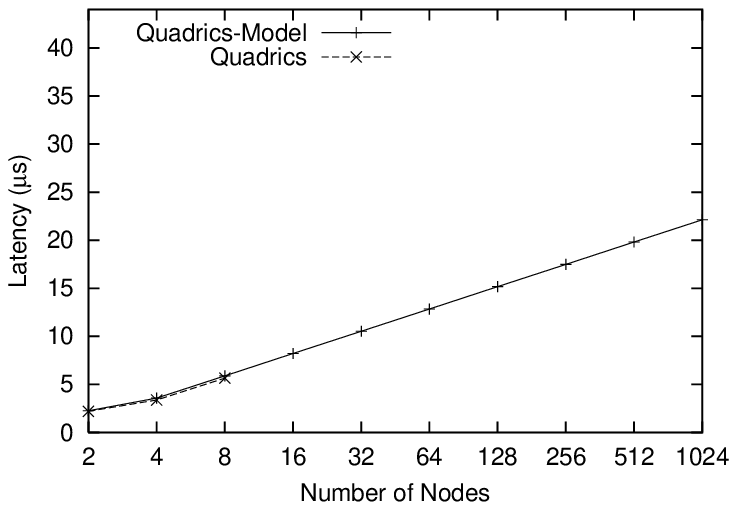}
    \label{fig:scale_q}
  }
  \hfill%
  \subfigure [Over 2.4MHz Myrinet-LANai-XP Cluster] {
    \includegraphics[width=0.47\columnwidth]{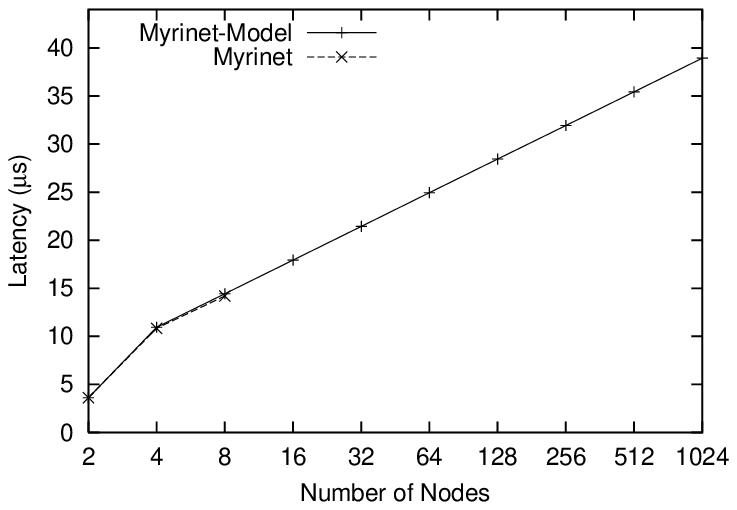}
    \label{fig:scale_m}
  }
  \vspace*{-0.5ex}
  \caption{Modeling of the Barrier Scalability}
  \label{fig:scalability}
\end{figure}

\SubSection{Scalability of the Proposed NIC-Based Barrier}
\label{SubSec:scalability-results}
As the size of parallel system reaches thousands,
it is important for parallel applications to be able to run over
larger size systems and achieve corresponding parallel speedup.
This requires that the underlying programming models provide scalable
communication, in particular, scalable collective operations.
Thus it is important to find out how the NIC-based barrier operations 
can scale over larger size systems.

Since the NIC-based barrier operations with 
the dissemination algorithm exhibits a consistent 
behavior as the system size increases, we choose its performance pattern
to model the scalability over different size systems. 
We formulate the latency for NIC-based barrier with the following equation.
\begin{displaymath}
T_{barrier} = T_{init} + (\lceil log_2 \: N \rceil - 1 ) * T_{trig} + T_{adj}
\end{displaymath}
In this equation, $T_{init}$ is the average NIC-based barrier latency
over two nodes, where each NIC only sends an initial barrier message 
for the entire barrier operation;
$T_{trig}$ is the average time for every other message 
the NIC needs to trigger when having received an earlier message;
and $T_{adj}$ is provided as the adjustment factor. 
The adjustment factor is needed to reflect the effects from other
aspects of the NIC-based barrier, e.g., 
reduced PCI bus traffic and the overhead of bookkeeping.
Through mathematical analysis,
we have derived Myrinet NIC-based barrier latency as 
$T_{barrier} = 3.60 + (\lceil log_2 \: N \rceil - 1 ) * 3.50 + 3.84$
for 2.4GHz Xeon clusters with LANai-XP cards, and
Quadrics NIC-based barrier latency as 
$T_{barrier} = 2.25 + (\lceil log_2 \: N \rceil - 1 ) * 2.32 - 1.00$
for quad-700MHz clusters with Elan3 cards.
As shown in Fig.~\ref{fig:scalability}, 
the NIC-based barrier operations could achieve
a barrier latency of  22.13$\mu$s and 38.94$\mu$s 
over a 1024-node Quadrics and Myrinet cluster of the same kinds,
respectively.
In addition, it indicates that the NIC-based barrier has potential 
for developing high performance
and scalable communication subsystems for next generation clusters.

\Section{Conclusions and Future Work}
\label{Sec:discussion}

We have characterized general concepts and the benefits of the
NIC-based barrier algorithms on top of point-to-point communication. 
We have then examined the communication processing 
for point-to-point operations, and pinpointed the relevant processing 
we can reduce for collective operations.
Accordingly we have proposed a general scheme 
for an efficient NIC-based barrier operation.  
The proposed scheme has been implemented over both Quadrics and Myrinet. 

Our evaluation has shown that, over a Quadrics cluster of 8 nodes with 
ELan3 Network, the NIC-based barrier operation 
achieves a barrier latency of only 5.60$\mu$s. This result is a 2.48 
factor of improvement over the Elanlib barrier operation 
when Quadrics hardware-based broadcast is not available.
In addition, our evaluation has also shown that, 
over a 16-node Myrinet cluster 
with LANai 9.1 cards, the NIC-based barrier operation achieves
a barrier latency of 25.72us, which is a 3.38 factor of 
improvement compared to the host-based algorithm.
Furthermore, our analytical model suggests that 
NIC-based barrier operations 
could achieve a latency of 22.13$\mu$s and 38.94$\mu$s,
respectively over a 1024-node Quadrics and Myrinet cluster.

As QsNet-II~\cite{Beecroft:qsnet_02}
and newer Myrinet interface cards become available to us, 
we are planning to investigate how this NIC-based barrier algorithm 
can accommodate and benefit from novel interconnect features.
In future, we also intend to study the benefits of this NIC-based
barrier for different parallel programming models and
applications built on top of them. 
Specifically, we plan to incorporate this barrier algorithm 
into LA-MPI~\cite{Graham:Network_ics_02_lampi}
to provide a more efficient barrier operation.
In addition, we intend to incorporate this NIC-based barrier, along with 
the NIC-based broadcast~\cite{Yuw:High_icpp_03_bcast}
into a resource management framework 
(e.g., STORM~\cite{Frachtenberg:storm_sc2002}) to 
investigate their benefits in increasing the resource utilization and 
the efficiency of resource management.
Furthermore, we intend to investigate whether other
collective communication operations, such as Allgather or Alltoall
could benefit from similar NIC-level implementations.


\begin{spacing}{0.9}
\bibliographystyle{latex8}
\bibliography{paper,panda4}                    

\begin{thebibliography}{10}\setlength{\itemsep}{-1ex}\small

\bibitem{Beecroft:qsnet_02}
J.~Beecroft, D.~Addison, F.~Petrini, and M.~McLaren.
\newblock {QsNet-II: An Interconnect for Supercomputing Applications}.
\newblock In {\em the Proceedings of Hot Chips '03}, Stanford, CA, August 2003.

\bibitem{Boden:Myrinet_95}
N.~J. Boden, D.~Cohen, R.~E. Felderman, A.~E. Kulawik, C.~L. Seitz, J.~N.
  Seizovic, and W.-K. Su.
\newblock {Myrinet: A Gigabit-per-Second Local Area Network}.
\newblock {\em IEEE Micro}, 15(1):29--36, 1995.

\bibitem{Buntinas:barrier_01}
D.~Buntinas, D.~K. Panda, and P.~Sadayappan.
\newblock {Fast NIC-Level Barrier over Myrinet/GM}.
\newblock In {\em IPDPS}, 2001.

\bibitem{Buntinas:Performance_cac_01_barrier}
D.~Buntinas, D.~K. Panda, and P.~Sadayappan.
\newblock Performance benefits of {NIC}-based barrier on {Myrinet/GM}.
\newblock In {\em CAC '01 Workshop (in conjunction with IPDPS)}, April 2001.

\bibitem{Eddington:infinibridge_02}
C.~Eddington.
\newblock {InfiniBridge: An InfiniBand Channel Adapter With Integrated Switch}.
\newblock {\em IEEE Micro}, (2):48--56, April 2002.

\bibitem{Frachtenberg:storm_sc2002}
E.~Frachtenberg, F.~Petrini, J.~Fernandez, S.~Pakin, and S.~Coll.
\newblock {STORM: Lightning-Fast Resource Management}.
\newblock In {\em Proceedings of the Supercomputing '02}, Baltimore, MD,
  November 2002.

\bibitem{Graham:Network_ics_02_lampi}
R.~L. Graham, S.-E. Choi, D.~J. Daniel, N.~N. Desai, R.~Minnich, C.~E.
  Rasmussen, {L. Dean Risinger}, and M.~W. Sukalski.
\newblock {A Network-Failure-tolerant Message-Passing system for Terascale
  Clusters}.
\newblock In {\em {ICS '02}}, June 2002.

\bibitem{Gropp:High_96_mpich}
W.~Gropp, E.~Lusk, N.~Doss, and A.~Skjellum.
\newblock {A High-Performance, Portable Implementation of the {MPI} Message
  Passing Interface Standard}.
\newblock {\em Parallel Computing}, 22(6):789--828, 1996.

\bibitem{Kinis:Fast_euro_20003_barrier}
S.~P. Kini, J.~Liu, J.~Wu, P.~Wyckoff, and D.~K. Panda.
\newblock {Fast and Scalable Barrier using RDMA and Multicast Mechanisms for
  InfiniBand-Based Clusters}.
\newblock In {\em {Euro PVM/MPI Conference}}, Venice, Italy, September 2003.

\bibitem{Liuj:Micro_hoti_03}
J.~Liu, B.~Chandrasekaran, W.~Yu, J.~Wu, D.~Buntinas, S.~Kinis, P.~Wyckoff, and
  D.~K. Panda.
\newblock {Micro-Benchmark Level Performance Comparison of High-Speed Cluster
  Interconnects}.
\newblock In {\em {Hot Interconnects 11}, (HotI 2003)}, Stanford, CA, August
  2003.

\bibitem{McKinley:Survey_94_collective}
P.~K. McKinley, Y.-J. Tsai, and D.~F. Robinson.
\newblock {A Survey of Collective Communication in Wormhole-Routed Massively
  Parallel Computers}.
\newblock Technical Report MSU-CPS-94-35, Dept. of Computer Science, Michigan
  State University, 1994.

\bibitem{Crummey:Algorithms_1991_ACM_sync}
J.~M. Mellor-Crummey and M.~L. Scott.
\newblock Algorithms for scalable synchronization on shared-memory
  multiprocessors.
\newblock {\em ACM Transactions on Computer Systems}, 9(1):21--65, 1991.

\bibitem{MPI:mpi_2_96}
{Message Passing Interface Forum, {MPIF}}.
\newblock {MPI-2: Extensions to the Message-Passing Interface}.
\newblock Technical Report, University of Tennessee, Knoxville, 1996.

\bibitem{Moody:Scalable_sc_03_reduce}
A.~Moody, J.~Fernandez, F.~Petrini, and D.~Panda.
\newblock {Scalable NIC-based reduction on Large-scale Clusters}.
\newblock In {\em {SC '03}}, Phoenix, Arizona, November 2003.

\bibitem{Petrini:Quadrics_micro_02_journal}
F.~Petrini, W.~chun Feng, A.~Hoisie, S.~Coll, and E.~Frachtenberg.
\newblock {The Quadrics Network: High Performance Clustering Technology}.
\newblock {\em IEEE Micro}, 22(1):46--57, January-February 2002.

\bibitem{Petrini:Hardware_nca_01_coll}
F.~Petrini, S.~Coll, E.~Frachtenberg, and A.~Hoisie.
\newblock {Hardware- and Software-Based Collective Communication on the
  Quadrics Network}.
\newblock In {\em NCA 2001}, Boston, MA, February 2002.

\bibitem{Quadrics:Web}
{Quadrics Supercomputers World, Ltd.}
\newblock {Quadrics Documentation Collection}.
\newblock {http://www.\linebreak[0]quadrics.com/}.

\bibitem{Yuw:High_icpp_03_bcast}
W.~Yu, D.~Buntinas, and D.~K. Panda.
\newblock {High Performance and Reliable NIC-Based Multicast over
  Myrinet/GM-2}.
\newblock In {\em {Int'l Conference on Parallel Processing}, (ICPP '03)},
  Kaohsiung, Taiwan, October 2003.

\end{thebibliography}
\end{spacing}


\clearpage

\begin{figure*}

The submitted manuscript has been created in part by the University of
Chicago as Operator of Argonne National Laboratory ("Argonne") under
Contract No.  W-31-109-ENG-38 with the U.S. Department of Energy.  The
U.S. Government retains for itself, and others acting on its behalf, a
paid-up, nonexclusive, irrevocable worldwide license in said article
to reproduce, prepare derivative works, distribute copies to the
public, and perform publicly and display publicly, by or on behalf of
the Government.
\thispagestyle{empty}

\end{figure*}

\end{document}